\documentclass[%
aps,prl,superscriptaddress,twocolumn,showpacs,preprintnumbers,amsmath,amssymb
]{revtex4-2}

\usepackage{graphicx}
\usepackage{color}
\usepackage{bm}
\usepackage{hyperref}
\usepackage{mathcomp}

\usepackage{amsmath}
\usepackage{amssymb}
\usepackage{empheq}
\usepackage{float}
\usepackage{multirow}
\usepackage{cases}
\usepackage{mathrsfs}
\usepackage{braket}

\begin{document}

%\preprint{APS/123-QED}

\title{Zero-field superconducting diode effect induced by magnetic flux in a van der Waals superconductor trigonal PtBi$_2$}% Force line breaks with \\

\author{Nan~Jiang}
\email{nan.jiang@phys.sci.osaka-u.ac.jp}
\affiliation{Department of Physics, Graduate School of Science, The University of Osaka, Toyonaka, Osaka 560-0043, Japan}
\affiliation{Center for Spintronics Research Network, The University of Osaka, Toyonaka 560-8531, Japan}
\affiliation{Institute for Open and Transdisciplinary Research Initiatives, The University of Osaka, Suita 565-0871,Japan}
\author{Masaki~Maeda}
\affiliation{Department of Physics, Graduate School of Science, The University of Osaka, Toyonaka, Osaka 560-0043, Japan}
\author{Yuhi~Yamaguchi}
\affiliation{Department of Physics, Graduate School of Science, The University of Osaka, Toyonaka, Osaka 560-0043, Japan}
\author{Mori~Watanabe}
\affiliation{Department of Physics, Graduate School of Science, The University of Osaka, Toyonaka, Osaka 560-0043, Japan}
\author{Masashi~Tokuda}
\affiliation{Department of Physics, Graduate School of Science, The University of Osaka, Toyonaka, Osaka 560-0043, Japan}
\author{Kensuke~Takaki}
\affiliation{Department of Physics, Graduate School of Science, The University of Osaka, Toyonaka, Osaka 560-0043, Japan}
\author{Sebun~Masaki}
\affiliation{Department of Physics, Graduate School of Science, The University of Osaka, Toyonaka, Osaka 560-0043, Japan}
\author{Takumi~Ikushima}
\affiliation{Department of Physics, Graduate School of Science, The University of Osaka, Toyonaka, Osaka 560-0043, Japan}
\author{Takayoshi~Koyanagi}
\affiliation{Department of Physics, Graduate School of Science, Tohoku University, Sendai 980-8578, Japan}
\author{Masakazu~Matsubara}
\affiliation{Center for Spintronics Research Network, The University of Osaka, Toyonaka 560-8531, Japan}
\affiliation{Institute for Open and Transdisciplinary Research Initiatives, The University of Osaka, Suita 565-0871,Japan}
\affiliation{Department of Physics, Graduate School of Science, Tohoku University, Sendai 980-8578, Japan}
\affiliation{Division of Materials Physics, Graduate School of Engineering Science, The University of Osaka, Toyonaka, Osaka 560-0043, Japan}
\affiliation{PRESTO, Japan Science and Technology Agency (JST), Kawaguchi 332-0012, Japan}
\author{Kazutaka~Kudo}
\affiliation{Department of Physics, Graduate School of Science, The University of Osaka, Toyonaka, Osaka 560-0043, Japan}
\affiliation{Institute for Open and Transdisciplinary Research Initiatives, The University of Osaka, Suita 565-0871,Japan}
\author{Yasuhiro~Niimi}
\affiliation{Department of Physics, Graduate School of Science, The University of Osaka, Toyonaka, Osaka 560-0043, Japan}
\affiliation{Center for Spintronics Research Network, The University of Osaka, Toyonaka 560-8531, Japan}
\affiliation{Institute for Open and Transdisciplinary Research Initiatives, The University of Osaka, Suita 565-0871,Japan}

\date{\today}% It is always \today, today,

\begin{abstract}
The superconducting diode effect is one of the nonreciprocal transport phenomena, where the critical current depends on the current direction. This effect is typically realized in superconductors with broken spatial-inversion and time-reversal symmetry. To break the time-reversal symmetry, external magnetic fields are commonly used. Here, we demonstrate a sign-controllable superconducting diode effect under zero external magnetic field in a van der Waals superconductor trigonal PtBi$_2$. The sign of the zero-field superconducting diode effect is controlled by the poling magnetic field, that is a large magnetic field applied prior to measurements. This result indicates that trapped magnetic flux are responsible for breaking the time-reversal symmetry. Our findings highlight the crucial role of trapped magnetic flux in generating the superconducting diode effect and provide a general pathway for realizing a zero-field superconducting diode effect.
\end{abstract}

\maketitle

\section{Introduction}
Nonreciprocal transport phenomena refer to situations where the resistivity depends on the direction of the current. Such phenomena typically require the breaking of both spatial-inversion symmetry and time-reversal symmetry, and they have been observed in systems such as distorted Bi~\cite{Bi}, noncentrosymmetric semiconductors~\cite{BiTeBr}, noncentrosymmetric magnets~\cite{MnSi,CNS,MnP,MnAu2}, topological semimetals~\cite{ZrTe5,WTe22,WTe2}, topological insulators~\cite{TI1,TI2}, bilayer metal films~\cite{PtCo}, and noncentrosymmetric superconductors~\cite{MoS2,STO,PbTaSe2}.

Recently, the concept of nonreciprocal transport in superconductors has been termed the superconducting diode effect~\cite{Ando2020}, where the magnitude of the critical current depends on the direction of the current. Consequently, near the critical current, the system exhibits either superconducting state or normal state depending on the current direction. 
To break spatial-inversion symmetry, various approaches have been employed, including artificial tricolor superlattices~\cite{Ando2020}, noncentrosymmetric crystal structures~\cite{Bauriedl2022,Zhang2020}, bilayer systems~\cite{Gutfreund2023}, and Josephson junctions~\cite{JD1,JD2,JD3,Matsuo2023,Gupta2023,Kim2024,Baumgartner2022,Pal2022,PhysRevLett.129.267702}. The shape of the sample also plays a significant role in the spatial-inversion symmetry due to the effects of Meissner screening currents and/or surface barriers~\cite{PhysRevB.72.172508,PhysRevLett.131.027001}. Thus, careful consideration is required to study the intrinsic spatial-inversion symmetry breaking that may cause the superconducting diode effect~\cite{He_2022,PhysRevLett.128.177001,pnas.2119548119,PhysRevLett.128.037001,PhysRevB.106.205206}. 

On the other hand, time-reversal symmetry breaking is typically achieved by an external magnetic field. Recently, however, zero-field superconducting diode effects have been observed in systems such as superconductors proximitized with ferromagnets~\cite{Jeon2022,adma.202304083,Narita2022,Xiong2024}, Josephson diodes~\cite{Wu2022}, twisted trilayer graphene on WSe$_2$~\cite{Lin2022,Scammell_2022}, twisted cuprate superconductors~\cite{science.abl8371}, chiral molecule-TaS$_2$ hybrid superlattices~\cite{Wan2024}, kagome superconductors~\cite{Le2024}, and strained noncentrosymmetric superconductors~\cite{sciadv.ado1502}.
Understanding the mechanisms that determine the sign of the zero-field superconducting diode effect is crucial for uncovering the origin of the diode effect as well as for enabling its potential application as a fundamental component in superconducting circuits. Although the sign reversal of the zero-field superconducting diode effect has been demonstrated in several systems~\cite{Jeon2022,adma.202304083,Narita2022,Xiong2024,Lin2022,Scammell_2022}, the methods and underlying physics in these cases are not readily transferable to more general systems. Therefore, developing universal approaches to break time-reversal symmetry and to achieve sign-switchable zero-field superconducting diode effects is highly desirable.

Magnetic flux is a typical mechanism for breaking time-reversal symmetry in superconductors. While their role in nonreciprocal transport under external magnetic fields has been widely discussed, there has been no report on a zero-field superconducting diode effect induced by magnetic flux in systems without magnetic materials.
In order to achieve a zero-field superconducting diode effect utilizing magnetic flux, it is advantageous to use a superconductor that can strongly trap magnetic flux, thereby influencing its transport properties.

Here, we focus on a van der Waals (vdW) superconductor, trigonal PtBi$_2$~\cite{ptbi2_1,ptbi2_lmr1,ptbi2_lmr2,10.1063/5.0137604,ptbi2rashba,ptbi2,PtBi2Press,TP,Kuibarov2024}, as a promising platform to realize a zero-field superconducting diode effect utilizing magnetic flux. Trigonal PtBi$_2$ is a noncentrosymmetric vdW material with a space group of $P31m$ (Fig.~1(a))~\cite{ptbi2_1}, possessing polarity $P$ along the $c$-axis and a large Rashba splitting~\cite{ptbi2rashba}. Furthermore, Se-doped trigonal PtBi$_2$, which undergoes a structural phase transition from a polar to a non-polar structure~\cite{takaki}, exhibits characteristic magnetoresistance induced by trapped magnetic flux~\cite{samukawa}. This behavior in Pt(Bi$_{1-x}$Se$_x$)$_2$ suggests that trigonal PtBi$_2$ also has a potential to trap magnetic flux. The combination of its noncentrosymmetric crystal structure and its potential to trap magnetic flux makes trigonal PtBi$_2$ a promising candidate for observing a zero-field superconducting diode effect utilizing magnetic flux.

Using trigonal PtBi$_2$, we have observed a sign-controllable superconducting diode effect under zero external magnetic field. The sign of this zero-field superconducting diode effect can be switched by poling magnetic fields, which are large magnetic fields applied before the measurements. We attribute this zero-field superconducting diode effect to the breaking of time-reversal symmetry by trapped magnetic flux, even in the absence of an external magnetic field. Our observations highlight the critical role of magnetic flux in the study of the superconducting diode effect and reveal a novel mechanism for achieving a sign-controllable superconducting diode effect.

\section{experimental methods}

\begin{figure}
\includegraphics[width=85mm]{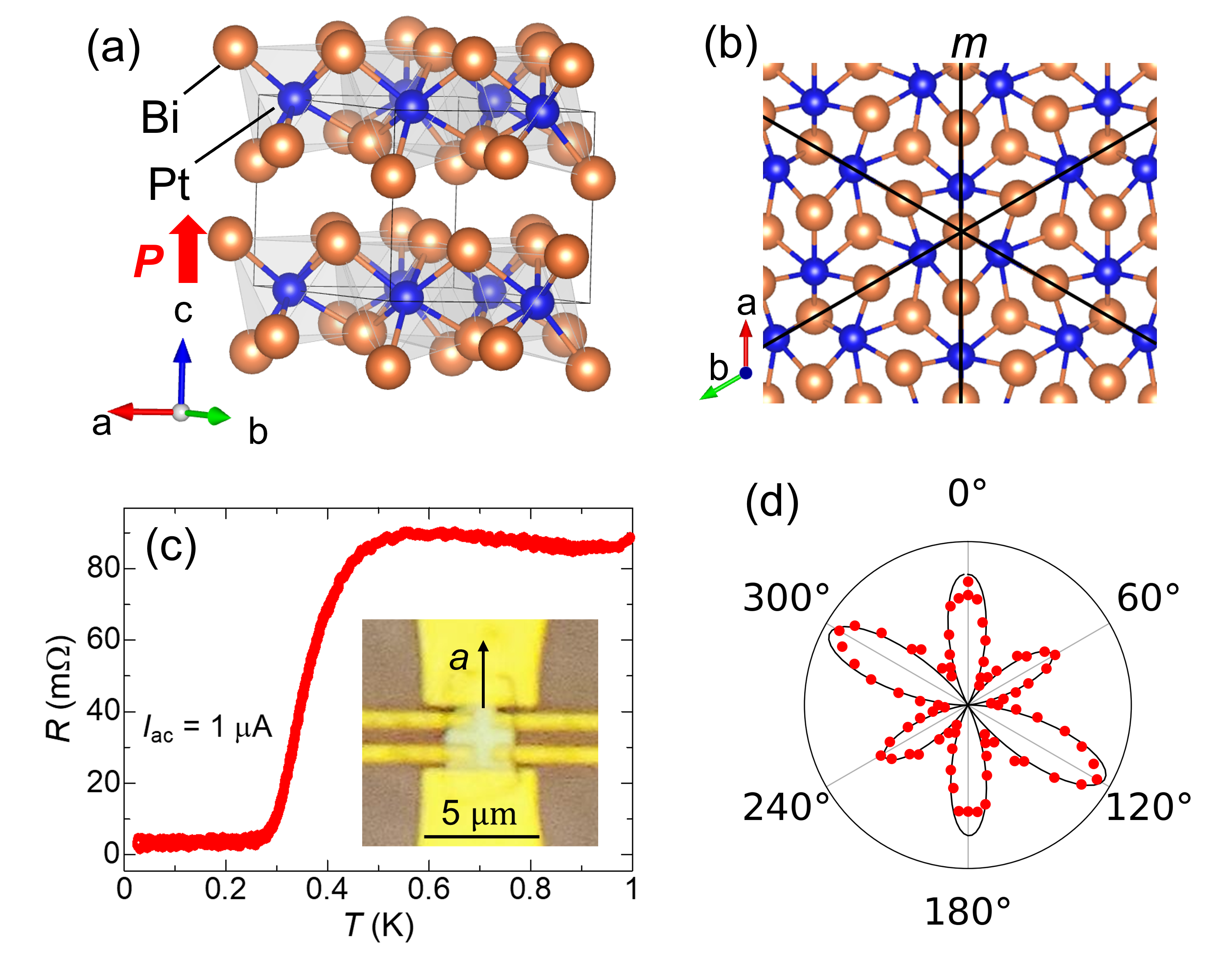}
\caption{\label{fig:epsart} (a) Crystal structure of trigonal PtBi$_2$. (b) Crystal structure of trigonal PtBi$_2$ viewed along the $c$-axis. The black lines represent the mirror planes. (c) Temperature dependence of resistance measured with a typical device (device A; thickness $95~\mathrm{nm}$). The inset shows an optical microscope image of device A. (d) Polarization analysis of the SHG signal for device A. The red circles represent the experimental data, while the black solid line denotes the fitting result. The observed six-lobe pattern corresponds to the mirror plane orientations. The angle $0^{\circ}$ is defined along the current flow direction (see Appendix~A for details).
}
\end{figure}

Single-crystalline bulk samples of trigonal PtBi$_2$ were synthesized using the same method as that for Pt(Bi$_{1-x}$Se$_x$)$_2$, previously reported by some of the present authors~\cite{takaki,samukawa}. Thin flakes of trigonal PtBi$_2$ with electrodes were prepared using a conventional lithography method, as shown in the inset of Fig.~1(c). Details of the fabrication process are provided in Appendix~A. 
The devices were cooled using a dilution refrigerator (Oxford Instruments) equipped with a superconducting magnet for transport measurements. Figure~1(c) shows the temperature dependence of resistance measured with a typical device, indicating a superconducting transition around 400~mK.

To investigate nonreciprocal transport phenomena in the vdW superconductor trigonal PtBi$_2$, we measured the differential resistance under DC electric currents to determine the critical currents as well as the first and second harmonic voltages under AC electric currents to observe the nonreciprocity, using the conventional four-terminal method. Measurements were performed with AC lock-in amplifiers (SR830 and LI5650) and preamplifiers (LI-75A and SA-440F5) at frequencies of 17~Hz or 173~Hz, in combination with a DC source (YOKOGAWA 7651). As discussed in previous studies~\cite{Bi,BiTeBr,MnSi,CNS,MnP,MnAu2,ZrTe5,WTe22,WTe2,TI1,TI2,PtCo,MoS2,STO,PbTaSe2}, the resistance of a system exhibiting nonreciprocity can be expressed as:
\begin{equation}
    R = R^{(0)} + R^{(1)} I,
\end{equation}
where $R^{(0)}$ and $R^{(1)}$ represent the normal and nonreciprocal resistances, respectively. 
When an AC current ($I = I_{0} \sin \omega t$) is applied, the output voltage can be written as:
\begin{align}
        V &= R^{(0)} I + R^{(1)} I^{2} \nonumber \\
      &= R^{(0)} I_{0} \sin \omega t + \frac{1}{2} R^{(1)} I_{0}^{2} \left( 1 + \sin \left( 2 \omega t - \frac{\pi}{2} \right) \right) \label{harmonics}.
\end{align}
Here, the first term in Eq.~(\ref{harmonics}) corresponds to the first harmonic voltage, and the second term in Eq.~(\ref{harmonics}) contains a DC offset and the second harmonic voltage. 
From these expressions, the first and second harmonic resistances are obtained as:
\begin{align}
    R^\omega &\equiv \frac{V^\omega}{I_0} = R^{(0)}, \\
    R^{2\omega} &\equiv \frac{V^{2\omega}}{I_0} = \frac{1}{2} R^{(1)} I_0.
\end{align}
Thus, by measuring the second harmonic resistance, we can quantify the contribution of nonreciprocal transport.

To determine the in-plane crystal orientation, we performed optical second harmonic generation (SHG) measurements. As described in Appendix~A, the mirror plane direction (see Fig.~1(b)) can be identified by analyzing the rotational anisotropy of the SHG intensity. The measurement was conducted at nearly normal incidence, detecting the SHG component $\theta_{2\omega}$ parallel to the incident polarization $\theta_{\omega}$ while rotating $\theta_{\omega} = \theta_{2\omega}$ through $360^{\circ}$ (see Fig.~7 for the experimental setup). The SHG intensity exhibited a distorted sixfold rotational symmetry (Fig.~1(d)), which is well reproduced by assuming $3m$ symmetry, consistent with the point group of trigonal PtBi$_{2}$. This result confirms that the current is applied parallel to the mirror plane.

\section{results}

\subsection{1. Nonreciprocal transport under an in-plane magnetic field}

\begin{figure}
\includegraphics[width=85mm]{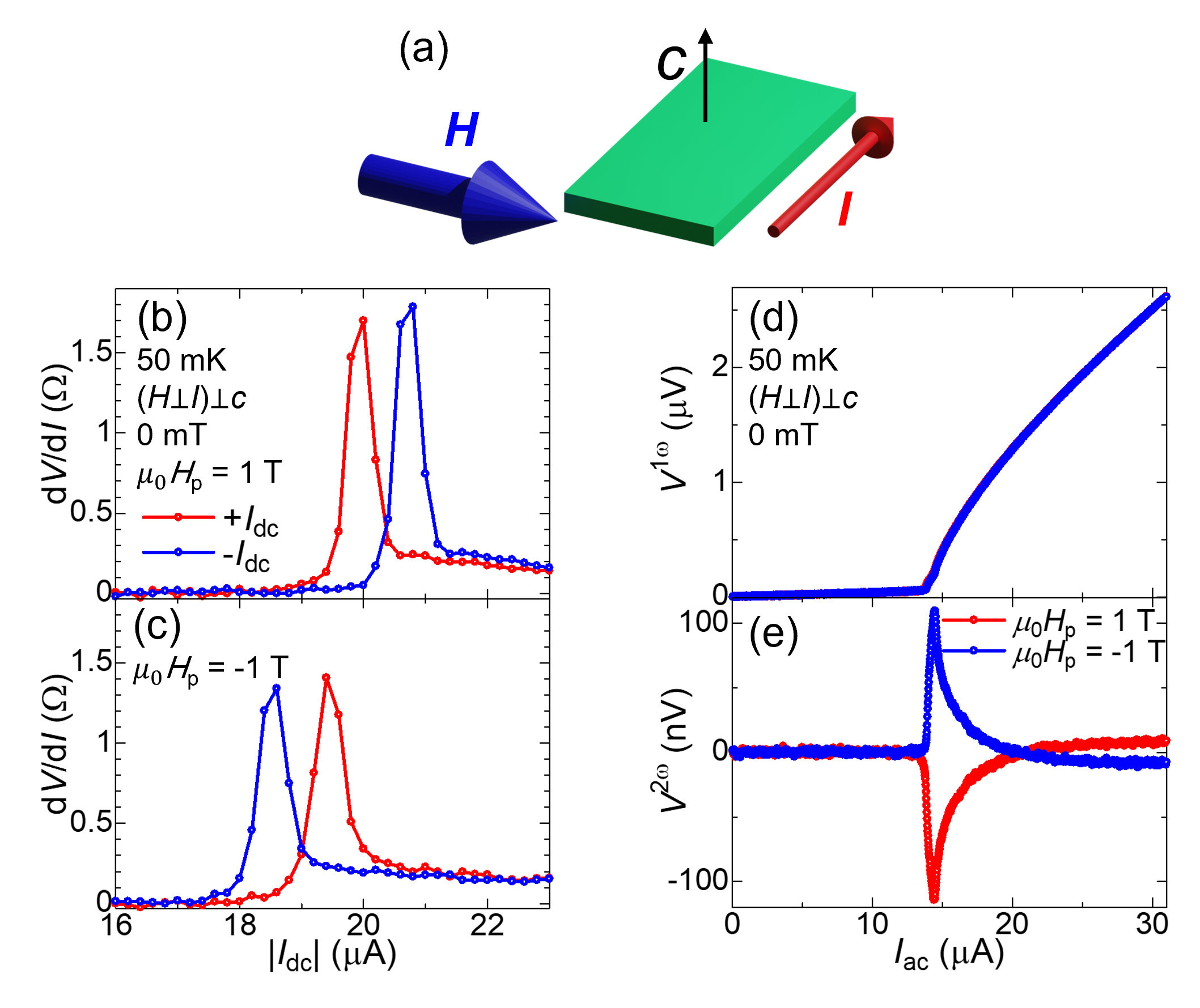}
\caption{\label{fig:epsart} (a) Schematic illustration of the measurement configuration under an in-plane magnetic field. 
(b), (c) $I_{\text{dc}}$ dependence of $\text{d}V/\text{d}I$ at 0~mT and 50~mK for (b) $\mu_0 H_{\rm p} = 1$~T and (c) $\mu_0 H_{\rm p} = -1$~T. Red and blue curves are measured with positive and negative $I_{\text{dc}}$. 
(d), (e) $I_{\text{ac}}$ dependence of (d) $V^{1\omega}$ and (e) $V^{2\omega}$ at 0~mT and 50~mK. Red and blue curves represent data for $\mu_0 H_{\rm p} = 1$~T and $\mu_0 H_{\rm p} = -1$~T, respectively.
}
\end{figure}

Firstly, we present nonreciprocal transport properties under an in-plane magnetic field perpendicular to the current, as illustrated in Fig.~2(a). To demonstrate the zero-field superconducting diode effect utilizing magnetic flux, we applied a large magnetic field ($\pm 1$~T) prior to the measurements, which is defined as a poling magnetic field $\mu_0 H_{\rm p}$. This field significantly exceeds the in-plane critical magnetic field ($\sim 75$~mT) and is used to control the trapped magnetic flux. The sign of the trapped magnetic flux can be tuned by changing the sign of $H_{\rm p}$.

Figure~2(b) shows the differential resistance $\text{d}V/\text{d}I$ as a function of the DC current $I_{\text{dc}}$ for the increasing process of $|I_{\text{dc}}|$, measured after applying $\mu_0 H_{\rm p} = 1$~T. 
The red and blue curves are plotted as a function of positive and negative $I_{\text{dc}}$, respectively. The critical current $I_{\rm c}$, defined as the $I_{\text{dc}}$ at which $\text{d}V/\text{d}I$ reaches its maximum, differs depending on the direction of $I_{\text{dc}}$, demonstrating a zero-field superconducting diode effect. 
When the sign of $H_{\rm p}$ is reversed from positive to negative, the sign of the zero-field superconducting diode effect also reverses, as shown in Fig.~2(c). These results confirm that the zero-field superconducting diode effect originates from trapped magnetic flux, whose sign is determined by the sign of $H_{\rm p}$.

The similar tendency has been observed in the second harmonic resistance measurements. Figures~2(d) and 2(e) show the AC current $I_{\text{ac}}$ (root mean square value) dependence of the first harmonic voltage $V^{1\omega}$ and the second harmonic voltage $V^{2\omega}$, respectively, obtained at $\mu_{0}H=0$~mT and $T=50$~mK. The red and blue curves represent data measured after applying a poling field of $\mu_0 H_{\rm p} = 1$~T and $\mu_0 H_{\rm p} = -1$~T, respectively. A pronounced $V^{2\omega}$ appears near the critical currents, indicating zero-field nonreciprocal transport phenomena. The sign of $V^{2\omega}$ also reverses with the sign of $H_{\rm p}$. 

\begin{figure}
\includegraphics[width=85mm]{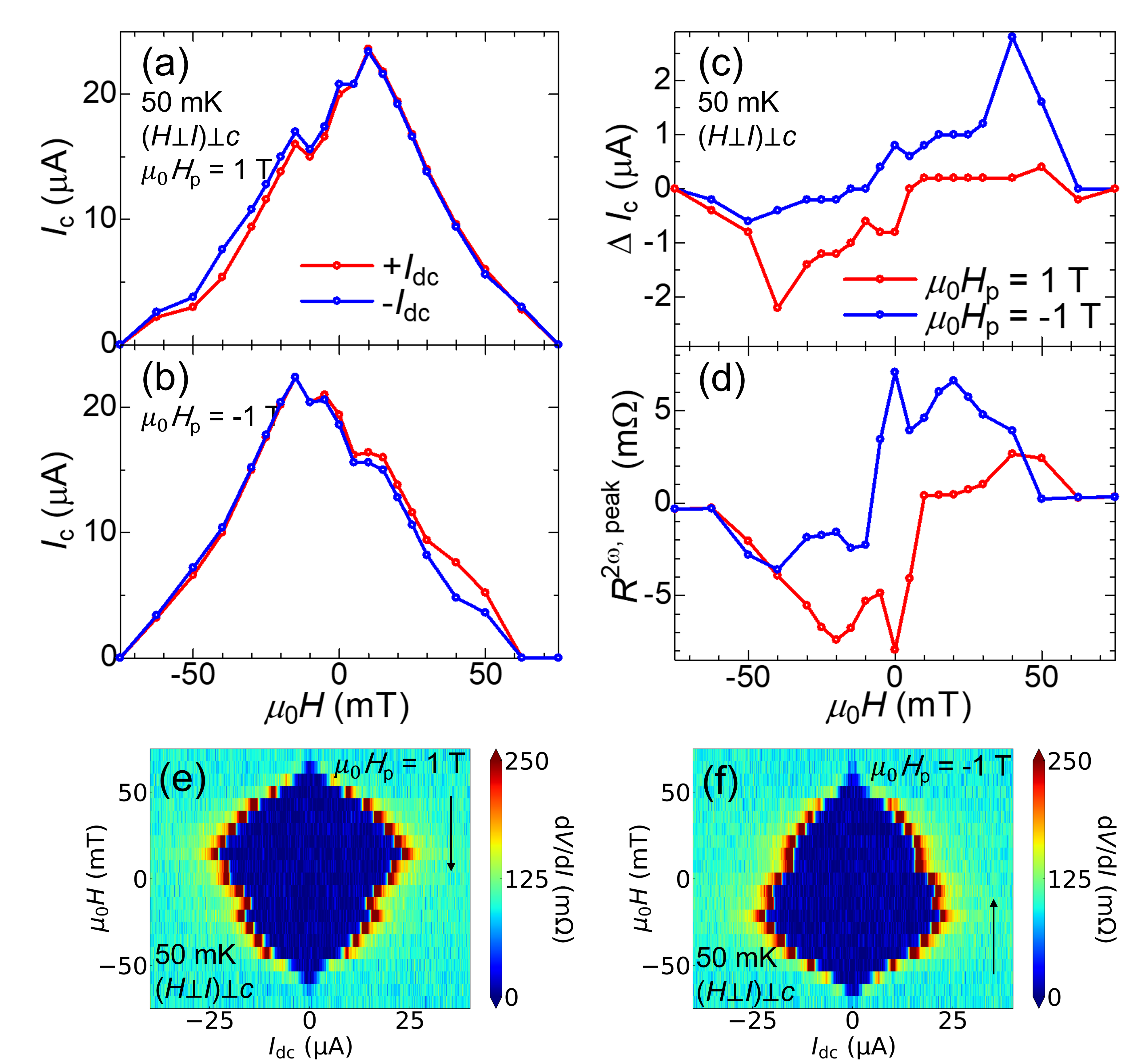}
\caption{\label{fig:epsart} (a), (b) In-plane magnetic field dependence of $I_{\text{c}}$ at 50~mK for (a) $\mu_0 H_{\rm p} = 1$~T and (b) $\mu_0 H_{\rm p} = -1$~T. Red and blue curves correspond to opposite $I_{\text{dc}}$ directions. 
(c), (d) In-plane magnetic field dependence of (c) $\Delta I_{\rm c}$ and (d) $R^{\rm 2\omega, peak}$ at 50~mK for $\mu_0 H_{\rm p} = 1$~T (red) and $\mu_0 H_{\rm p} = -1$~T (blue).
(e), (f) $\text{d}V/\text{d}I$ as a function of $I_{\text{dc}}$ and $\mu_0 H$ at 50~mK, measured after applying (e) $\mu_0 H_{\rm p} = 1$~T and (f) $\mu_0 H_{\rm p} = -1$~T.
}
\end{figure}

To examine the details of the nonreciprocal transport properties, we measured the in-plane magnetic field dependence. Figures~3(a) and 3(b) show the in-plane magnetic field dependence of $I_{\text{c}}$ at $T=50$~mK for $\mu_0 H_{\rm p} = 1$~T and $\mu_0 H_{\rm p} = -1$~T, respectively. Red and blue curves are measured with positive and negative $I_{\text{dc}}$, respectively. To evaluate the superconducting diode effect, we plot the difference in critical currents $\Delta I_{\rm c}$ between positive and negative $I_{\text{dc}}$ in Fig.~3(c) for both $\mu_0 H_{\rm p} = 1$~T (red) and $\mu_0 H_{\rm p} = -1$~T (blue). The superconducting diode effect depends on the in-plane magnetic field and the magnitude changes with respect to $H_{\rm p}$.

We have observed the similar behavior in the second harmonic resistance measurements. To evaluate the nonreciprocity, we plot the maximum of the second harmonic resistance $R^{\rm 2\omega, peak}$ as a function of the in-plane magnetic field in Fig.~3(d) for both $\mu_0 H_{\rm p} = 1$~T (red) and $\mu_0 H_{\rm p} = -1$~T (blue). $R^{\rm 2\omega, peak}$ also exhibits the same in-plane magnetic field dependence as $\Delta I_{\rm c}$. It should be noted that both $\Delta I_{\rm c}$ and $R^{\rm 2\omega, peak}$ retain non-zero values at zero magnetic field. The details of the magnetic field dependence of the nonreciprocity is discussed in Appendix~B.

The hysteretic behavior with respect to $H_{\rm p}$ appears not only in the nonreciprocal transport properties but also in the current-magnetic field phase diagram. Figures~3(e) and 3(f) show the color map of $\text{d}V/\text{d}I$ as a function of $I_{\text{dc}}$ and $\mu_0 H$ at $T=50$~mK for $\mu_0 H_{\rm p} = 1$~T and $\mu_0 H_{\rm p} = -1$~T, respectively. The blue region corresponds to the superconducting state with zero resistance, while the light green region indicates the normal resistance state. The area of zero resistance depends on the sign of $H_{\rm p}$. This hysteretic behavior supports the idea that trapped magnetic flux play a crucial role in the transport properties of trigonal PtBi$_2$ and that $H_{\rm p}$ can control the trapped magnetic flux. Transport properties for $|\mu_0 H_{\rm p}| = 2$~T, presented in Appendix~C, further support this scenario.

Such hysteresis can often arise from a stray magnetic field due to the superconducting magnet. To rule out this possibility, we measured the resistance of few-layer graphene simultaneously during all measurements. Since the magnetoresistance of graphene is an even function with respect to the external magnetic field and does not exhibit hysteresis, we used it to calibrate the external magnetic field. Details of this calibration are provided in Appendix~D. The external magnetic field $\mu_{0}H$ shown in the main text has already been corrected with this method.

\subsection{2. In-plane magnetic field direction dependence}

\begin{figure}
\includegraphics[width=85mm]{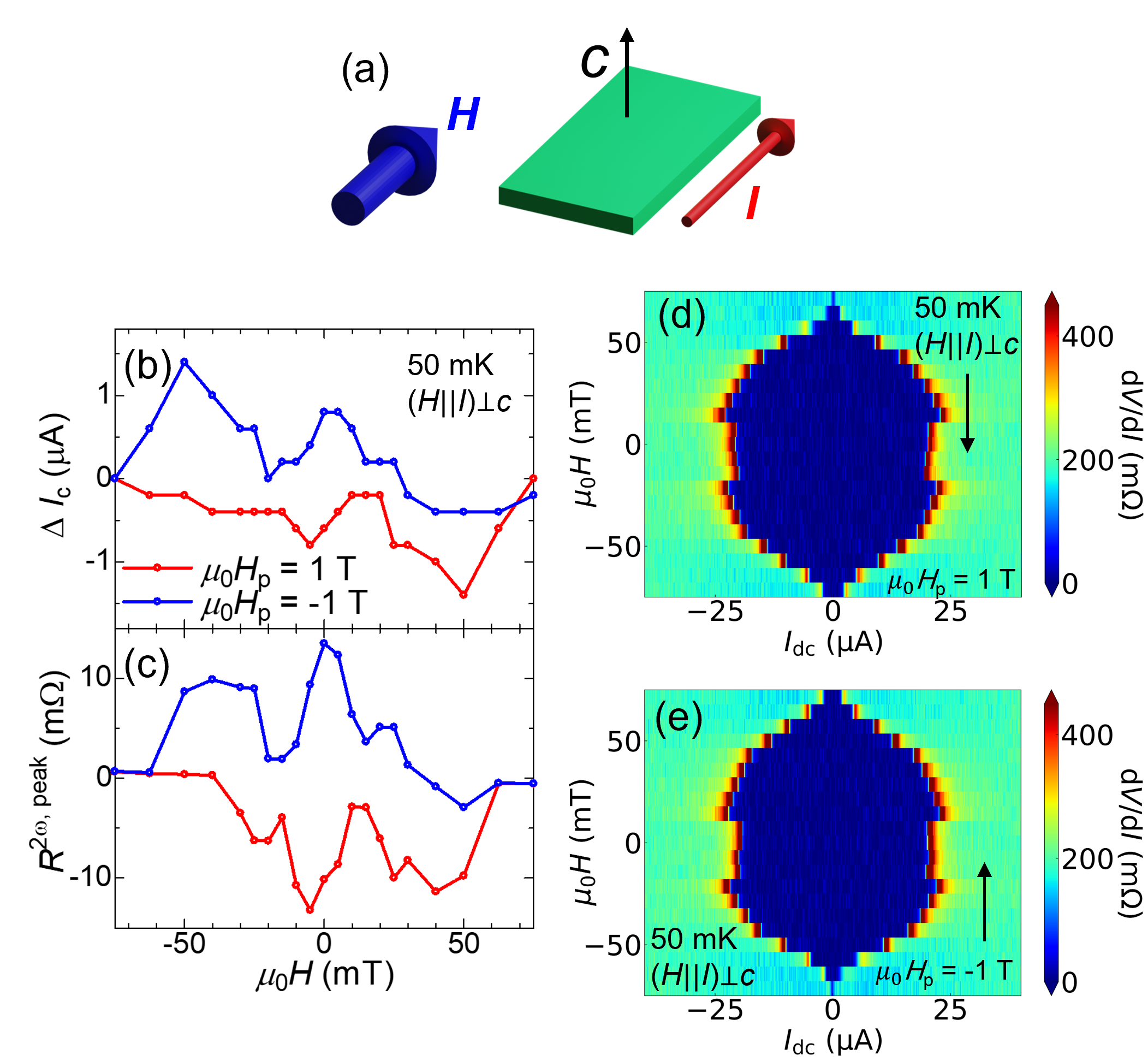}
\caption{\label{fig:epsart} (a) Schematic illustration of the measurement configuration under an in-plane magnetic field parallel to the current. 
(b), (c) $\Delta I_{\rm c}$ (b) and $R^{\rm 2\omega, peak}$ (c) at 50~mK for $\mu_0 H_{\rm p} = 1$~T (red) and $\mu_0 H_{\rm p} = -1$~T (blue) as a function of in-plane magnetic field parallel to the current. 
(d), (e) $\text{d}V/\text{d}I$ as a function of $I_{\text{dc}}$ and $\mu_0 H$ along the current direction, measured at 50~mK after applying (d) $\mu_0 H_{\rm p} = 1$~T and (e) $\mu_0 H_{\rm p} = -1$~T. 
}
\end{figure}

To investigate the origin of the nonreciprocal transport properties at zero magnetic field, we applied an in-plane magnetic field parallel to the current, as shown in Fig.~4(a). If this nonreciprocity originates from the polarity $P$ ($\parallel c$) of the crystal structure of trigonal PtBi$_2$, it should only appear in the $P \perp I \perp H$ configuration and not in the $P \perp (I \parallel H)$ configuration. However, we have observed nonreciprocal transport properties at zero magnetic field even in the $I \parallel H$ configuration. Figures~4(b) and 4(c) show $\Delta I_{\rm c}$ and $R^{\rm 2\omega, peak}$ at $T=50$~mK for both $\mu_0 H_{\rm p} = 1$~T (red) and $\mu_0 H_{\rm p} = -1$~T (blue), respectively, as a function of in-plane magnetic field parallel to the current. Nonreciprocal transport properties at zero magnetic field and a hysteresis with respect to $H_{\rm p}$ are clearly observed. 

The hysteresis with respect to $H_{\rm p}$ also appear in the current-magnetic field phase diagram.
Figures~4(d) and 4(e) show color maps of $\text{d}V/\text{d}I$ as a function of $I_{\text{dc}}$ and $\mu_0 H$ along the current direction, measured at $T=50$~mK after applying $\mu_0 H_{\rm p} = 1$~T and $\mu_0 H_{\rm p} = -1$~T, respectively. Although there are some differences between the $I \perp H$ configuration (Figs.~3(e) and 3(f)) and the $I \parallel H$ configuration (Figs.~4(d) and 4(e)), a hysteresis is also observed for the $I \parallel H$ configuration. In other words, the nonreciprocal transport can be seen not only in the $P \perp I \perp H$ configuration but also in the $P \perp (I \parallel H)$ configuration, particularly at zero magnetic field. These findings indicate that the nonreciprocal transport properties under an in-plane magnetic field are not solely caused by the polarity $P$ of trigonal PtBi$_2$. We will discuss the origin of the nonreciprocal transport properties again in the discussion section.

\subsection{3. Nonreciprocal transport under an out-of-plane magnetic field}

\begin{figure}
\includegraphics[width=85mm]{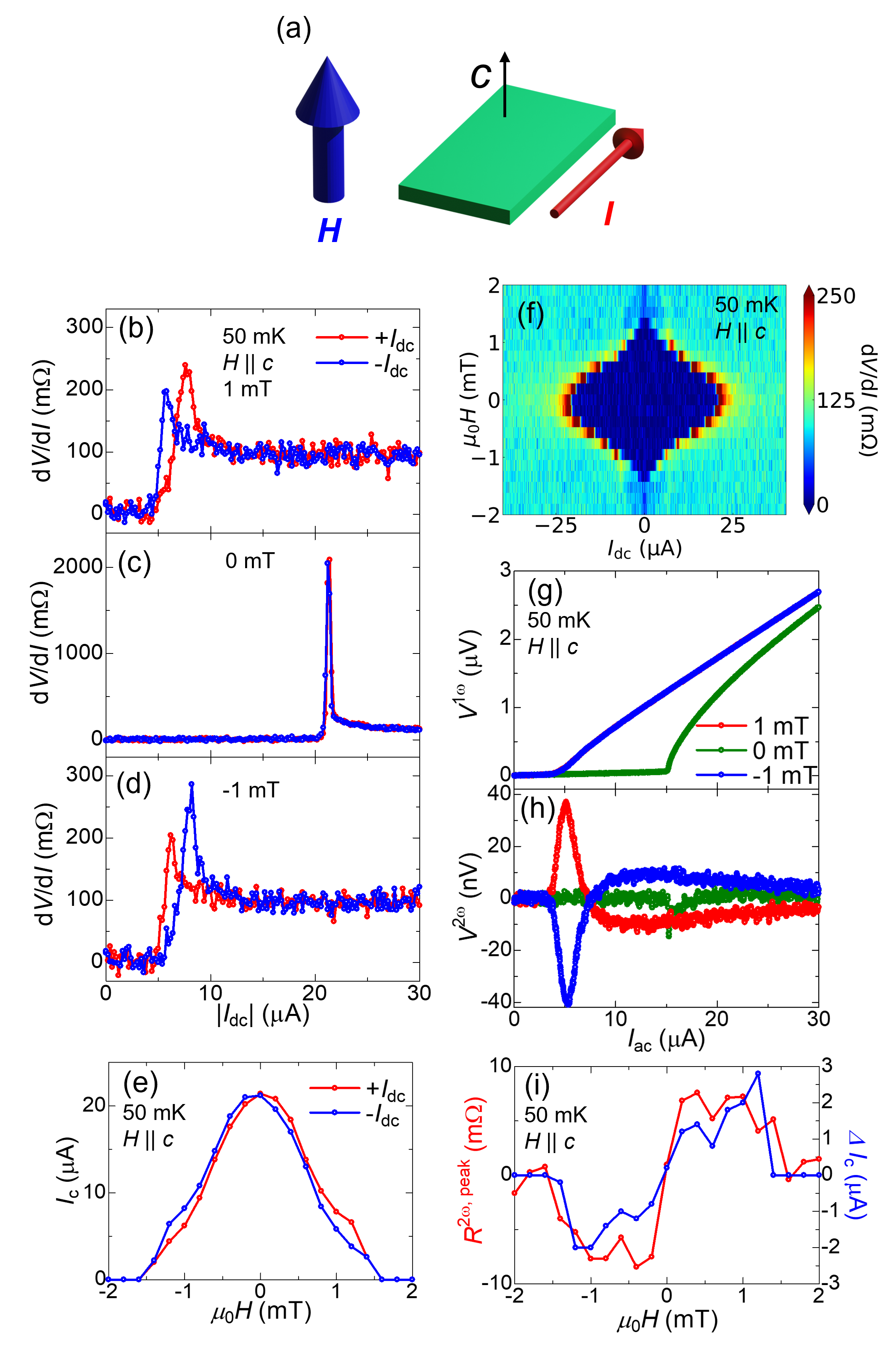}
\caption{\label{fig:epsart} (a) Schematic illustration of the measurement configuration under an out-of-plane magnetic field. 
(b)-(d) $\text{d}V/\text{d}I$ as a function of $I_{\text{dc}}$ under (b) $\mu_0 H = 1$, (c) 0, and (d) $-1$~mT at 50~mK. Red and blue curves correspond to opposite $I_{\text{dc}}$ directions, showing both increasing and decreasing processes of $I_{\text{dc}}$. 
(e) Out-of-plane magnetic field dependence of $I_{\rm c}$ at 50~mK. Red and blue curves are measured with positive and negative $I_{\text{dc}}$, respectively. 
(f) $\text{d}V/\text{d}I$ as a function of $I_{\text{dc}}$ and $\mu_0 H$ at 50~mK, with $H \parallel c$. 
(g), (h) $I_{\text{ac}}$ dependence of (f) $V^{1\omega}$ and (g) $V^{2\omega}$ at 50~mK. Red, green and blue curves represent data under 1~mT, 0~mT and $-1$~mT, respectively. 
(i) Out-of-plane magnetic field dependence of $R^{\rm 2\omega, peak}$ and $\Delta I_{\rm c}$ at 50~mK.
}
\end{figure}

Finally, we investigate the nonreciprocal transport properties under an out-of-plane magnetic field aligned parallel to the $c$-axis, as illustrated in Fig.~5(a). Figures~5(b)–5(d) show the $I_{\text{dc}}$ dependence of $\text{d}V/\text{d}I$ at 50~mK under (b) $\mu_0 H = 1$, (c) 0, and (d) $-1$~mT, respectively. The red and blue curves are obtained with the positive and negative $I_{\text{dc}}$, respectively. $I_{\rm c}$ depends on the sign of $I_{\text{dc}}$ under a finite out-of-plane magnetic field, indicating the presence of a superconducting diode effect. The sign of this diode effect under the out-of-plane magnetic field reverses with the magnetic field direction (Figs.~5(b) and 5(d)) and vanishes at zero magnetic field (Fig.~5(c)). To examine this behavior in more detail, Fig.~5(e) shows the out-of-plane magnetic field dependence of $I_{\rm c}$ at $T=50$~mK, where the red and blue curves are obtained with positive and negative $I_{\text{dc}}$, respectively. A color map of $\text{d}V/\text{d}I$ as a function of $I_{\text{dc}}$ and $\mu_0 H$ at $T=50$~mK with $H \parallel c$ is shown in Fig.~5(f). The field dependence of the superconducting diode effect, characterized by $\Delta I_{\rm c}$, is presented in Fig.~5(i) (see blue data), showing an odd function with respect to the out-of-plane magnetic field.

The similar tendency has also been observed in second harmonic resistance measurements. Figures~5(g) and 5(h) show the $I_{\text{ac}}$ dependence of $V^{1\omega}$ and $V^{2\omega}$ at $T=50$~mK, respectively. The red, green, and blue data correspond to measurements under $\mu_0 H = 1$, 0, and $-1$~mT, respectively. Near the critical current, a significant nonreciprocity is detected in $V^{2\omega}$ under an out-of-plane magnetic field (red and blue data in Fig.~5(h)), which disappears at zero magnetic field (green data in Fig.~5(h)). The sign of $V^{2\omega}$ reverses with the magnetic field direction. To investigate this behavior further, Fig.~5(i) shows the out-of-plane magnetic field dependence of $R^{\rm 2\omega, peak}$ (see red data), which also exhibits an odd function with respect to the magnetic field.

As shown in Fig.~5(i), both $R^{\rm 2\omega, peak}$ and $\Delta I_{\rm c}$ are asymmetric with respect to the out-of-plane magnetic field and become zero at zero magnetic field. Moreover, the nonreciprocity under the out-of-plane magnetic field does not display hysteretic behavior with respect to $H_{\rm p}$, as demonstrated in Appendix~E. These results confirm that the zero-field nonreciprocal transport properties and the hysteresis with respect to $H_{\rm p}$ are unique to in-plane magnetic fields.

\section{Discussions}
In this section, we discuss the origin of the nonreciprocal properties in trigonal PtBi$_2$. To generate nonreciprocity within the linear response regime, both time-reversal symmetry and spatial-inversion symmetry should be broken. In our case, regarding spatial-inversion symmetry breaking, contributions can arise not only from the noncentrosymmetric crystal structure but also from the asymmetry of the sample shape, as reported in other superconductors~\cite{PhysRevB.72.172508,PhysRevLett.131.027001}. Thus, it is challenging to determine which specific mechanism of spatial-inversion symmetry breaking mainly contributes to the nonreciprocal transport in trigonal PtBi$_2$ under both in-plane and out-of-plane magnetic field configurations. As shown in Figs.~2-4, we observe zero-field nonreciprocal transport in both the $I \perp H$ and $I \parallel H$ configurations. This indicates that the nonreciprocal transport under an in-plane magnetic field does not solely originate from the polarity $P$ of trigonal PtBi$_2$. Furthermore, we have examined the reproducibility of the nonreciprocal transport in flakes with the same polarity $P$, as discussed in Appendix~F, and observed opposite signs of the nonreciprocal transport at zero magnetic field, even in flakes with the same $P$. These findings further support the conclusion that the nonreciprocal transport under an in-plane magnetic field is not solely caused by the polarity $P$ of trigonal PtBi$_2$, but that the asymmetry of the sample shape also plays an important role.

For the out-of-plane magnetic field configuration, the shape of the sample edge plays a crucial role in generating the superconducting diode effect~\cite{PhysRevB.72.172508,PhysRevLett.131.027001}. In particular, achieving a symmetric sample shape is challenging in exfoliated flakes. Therefore, asymmetric surface barriers for magnetic flux arising from the nonuniform sample shape are likely the most plausible origin of the superconducting diode effect under the out-of-plane magnetic field in our devices.

On the other hand, regarding time-reversal symmetry breaking, because trigonal PtBi$_2$ does not exhibit magnetic ordering, trapped magnetic flux are responsible for breaking time-reversal symmetry, even at zero magnetic field. As shown in Appendix~E, neither nonreciprocity at zero magnetic field nor hysteresis with respect to $H_{\rm p}$ is observed under the out-of-plane magnetic field configuration, indicating that trigonal PtBi$_2$ can trap magnetic flux under an in-plane magnetic field but not under an out-of-plane field. This anisotropy can be attributed to the van der Waals nature of trigonal PtBi$_2$. Similar phenomena and anisotropic magnetic flux trapping have been reported in Se-doped trigonal PtBi$_2$~\cite{samukawa}, but further studies are required to elucidate the detailed mechanism of magnetic flux trapping in trigonal PtBi$_2$.

Before closing this section, we would like to mention the limitations of the present study. Our proposed explanation, that the time reversal symmetry breaking at zero magnetic field in trigonal PtBi$_2$ is caused by trapped magnetic flux, remains speculative at this point. To obtain direct evidence, microscopic imaging techniques such as scanning tunneling microscopy or magneto-optical Kerr microscopy would be useful. These measurements could also provide important insights into sample uniformity and its possible influence on the observed hysteresis and nonreciprocal response.
In addition, since it is difficult to fabricate perfectly symmetric exfoliated nanoflakes, future studies may benefit from using samples prepared by molecular beam epitaxy and patterned through lithographic processes to ensure better geometric symmetry. Such improvements would enable a more systematic investigation of the intrinsic spatial inversion symmetry breaking in trigonal PtBi$_2$, including the dependence of nonreciprocal transport on the direction of the in plane current.

\section{Conclusions}
In conclusion, we have demonstrated the zero-field superconducting diode effect in a vdW superconductor trigonal PtBi$_2$ induced by magnetic flux. The sign of the zero-field superconducting diode effect reverses when the poling magnetic field $H_{\rm p}$ is inverted, confirming that the effect originates from trapped magnetic flux, whose sign is controlled by $H_{\rm p}$. 
The zero-field superconducting diode effect and its hysteretic behavior with respect to $H_{\rm p}$ are observed exclusively under in-plane magnetic fields. This suggests that trigonal PtBi$_2$ can trap magnetic flux more effectively under in-plane magnetic fields than under out-of-plane fields.
Our findings highlight the critical role of trapped magnetic flux in generating the superconducting diode effect and provide a general pathway for realizing sign-switchable zero-field superconducting diode effects.

\section{data availability}
Data are available from the corresponding authors upon request.

\section{acknowledgements}
The authors thank A. Daido, Y. Yanase, H. Adachi, and D. Sekine for fruitful discussions. The crystal structure of trigonal PtBi$_2$ was visualized using VESTA~\cite{vesta}. This work was supported by JSPS KAKENHI (Grant Nos. JP22H04481, JP22H01182, JP23H00257, JP23H04868, JP23K13062, JP23K22453, JP24H00413, JP24H01639, and JP24K21531), JST FOREST (Grant No. JPMJFR2134), JST PRESTO (Grant No. JPMJPR23H9), JST CREST (Grant No. JPMJCR24R5), the Asahi Glass Foundation, the TEPCO Memorial Foundation, TIA-Kakehashi (Grant No. TK24-13).

\section{Author contributions}
N.J. performed the device fabrication and transport measurements with help from M.M., Y.Y., M.W., and M.T.. K.T., S.M., and K.K. carried out the crystal growth. N.J., T.K., and M.M. conducted the SHG measurements. N.J. wrote the manuscript with input from Y.N..

\section{appendix A: experimental details}

\subsection{1. Device fabrication}

\begin{figure}
\includegraphics[width=85mm]{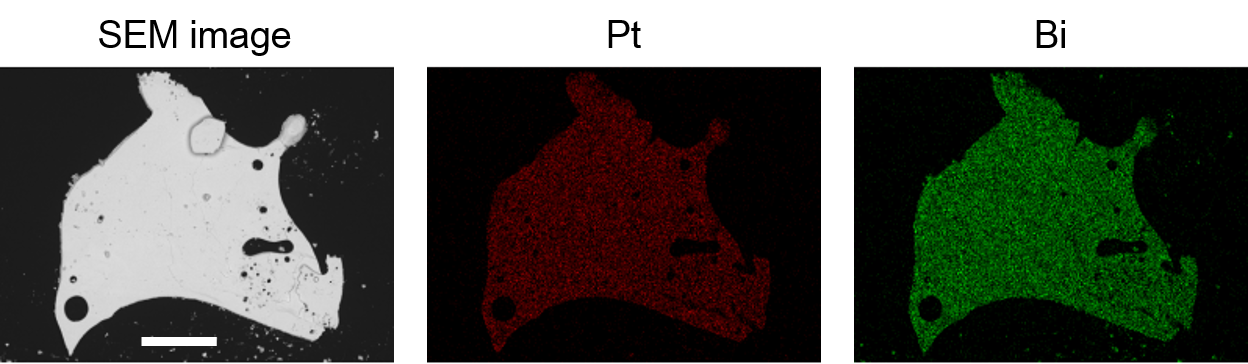}
\caption{\label{fig:epsart} Scanning Electron Microscope (SEM) image and EDS elemental maps of PtBi$_2$. Each represents the elemental mapping of Pt(red) and Bi(green). The white bars correspond to 100 $\mu$m.
}
\end{figure}

Single-crystalline bulk samples of trigonal PtBi$_2$ were synthesized using the same method as that for Pt(Bi$_{1-x}$Se$_x$)$_2$, previously reported by some of the present authors~\cite{takaki,samukawa}. A stoichiometric mixture of Pt (99.95\%) and Bi (99.99\%) powders was sealed in an evacuated quartz tube, heated at 630~$^{\circ}$C for 24 hours, and subsequently quenched in ice water. 
The chemical compositions of the synthesized samples were analyzed using energy dispersive X-ray spectroscopy (EDS) with a TM4000Plus II scanning electron microscope (Hitachi High-Tech) equipped with an AztecOne energy dispersive spectrometer (Oxford Instruments), as shown in Fig.~6. The sample composition is spatially uniform within a spatial resolution of approximately 2-3 $\mu$m. The crystal structure was examined by powder X-ray diffraction (XRD) using a MiniFlex600-C X-ray diffractometer with a D/teX Ultra2 high-speed one-dimensional detector (Rigaku).

Thin flakes of trigonal PtBi$_2$ were obtained by mechanical exfoliation using scotch tape. The exfoliated flakes were transferred onto Si/SiO$_2$ substrates, and electrodes were patterned by electron beam lithography, followed by the deposition of Ti/Au films. The exfoliation process was conducted inside a glove box filled with Ar gas of 99.9999\% purity to prevent oxidation of trigonal PtBi$_2$. An optical microscope image of a typical device is shown in the inset of Fig.~1(c). The thickness $d$ of each flake was measured using an atomic force microscope (NanoNaviReal Probe Station). 
The devices were cooled using a dilution refrigerator (Oxford Instruments) equipped with a superconducting magnet for transport measurements.

\subsection{2. Optical SHG measurement for crystal orientation determination}

\begin{figure}
	\begin{center}
	\includegraphics[width=0.7\linewidth]{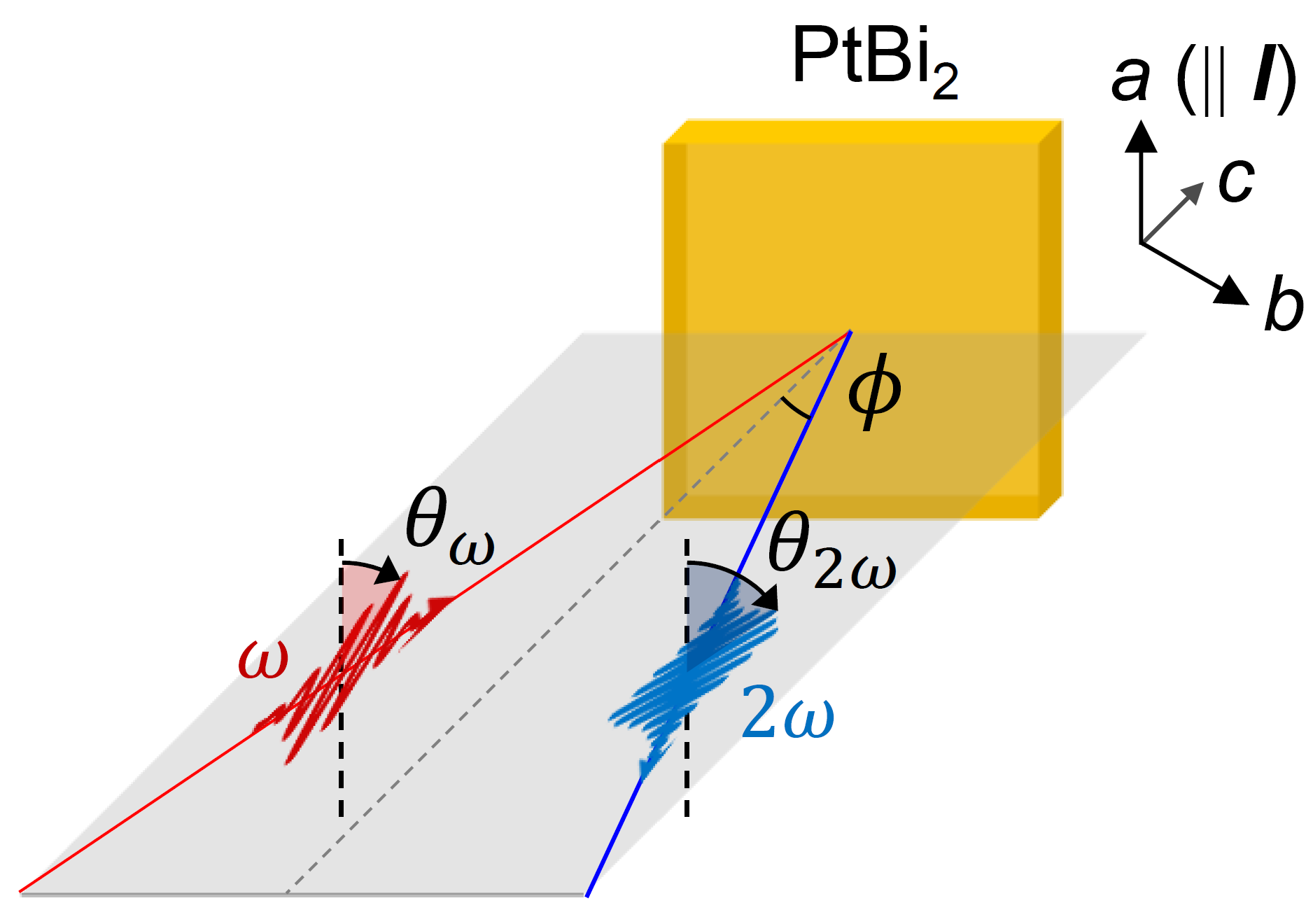}
	\end{center}
	\caption{Schematic illustration of the SHG measurement setup at an incident angle $\phi$. The angles $\theta_{\omega}$ and $\theta_{2\omega}$ denote the polarization directions of the incident fundamental light $(\omega)$ and the emitted SHG light ($2\omega$), respectively.}
	\label{fig:SHG_sup}
\end{figure}

To determine the in-plane crystal orientation of trigonal PtBi$_{2}$, we conducted optical second harmonic generation (SHG) measurements~\cite{Matsubara_2015_Science, Matsubara_2010_PRB}.
SHG is a nonlinear optical process in which two photons at frequency of $\omega$ interact with a material, generating a photon at $2\omega$.
Under the electric dipole (ED) approximation, SHG is given by:
\begin{equation}
P_i(2\omega) = \varepsilon_0 \sum_{j, k} \chi_{ijk}^{(2)} E_j(\omega) E_k(\omega)
\end{equation}
where $\varepsilon_0$ is the vacuum permittivity, $E_j(\omega)$ and $E_k(\omega)$ are the $j$- and $k$-polarized incident electric fields, and $P_i(2\omega)$ is the induced $i$-polarized nonlinear polarization.
ED-type SHG is allowed only in noncentrosymmetric materials or at surfaces/interfaces where inversion symmetry is broken.
Since the nonlinear susceptibility tensor $\chi_{ijk}^{(2)}$ is highly sensitive to crystal symmetry, SHG serves as a powerful, non-destructive tool for crystallographic orientation analysis.

In our experiment, ultrafast laser pulses (800 nm wavelength, $<$55 fs pulse width, 20 kHz repetition rate) irradiated the trigonal PtBi$_{2}$ surface. The SHG response was measured in a reflection geometry at an incident angle of $\sim 6^\circ$.
All measurements were conducted at room temperature in ambient conditions.

Trigonal PtBi$_{2}$ crystallizes in the point group $3m$ at room temperature.
Taking the $z (c)$-axis as the threefold rotational axis, the nonzero components of $\chi_{ijk}^{(2)}$ are:
\begin{align}
    \chi_{zzz}, \chi_{zxx} &= \chi_{zyy}, \quad \chi_{xxz} = \chi_{xzx} = \chi_{yyz} = \chi_{yzy}, \notag \\
    \chi_{yyy} &= -\chi_{yxx} = -\chi_{xyx} = -\chi_{xxy}.
    \label{eq:chi_components}
\end{align}

At oblique incidence, the nonlinear polarization components are expressed as:
\begin{align}
    P_x(2\omega) &= -2\varepsilon_0 \big( \chi_{yyy} \sin{\theta_{\omega}} \cos{\theta_{\omega}} \cos{\phi}  \notag \\
    &\quad + \chi_{zxx} \sin^{2}{\theta_{\omega}} \sin{\phi} \cos{\phi} \big) E(\omega)^2,  \notag \\
    P_y(2\omega) &= \varepsilon_0 \big( \chi_{yyy} \cos^{2}{\theta_{\omega}} - \chi_{yyy} \sin^{2}{\theta_{\omega}} \cos^{2}{\phi}  \notag \\
    &\quad - 2 \chi_{xxz} \sin{\theta_{\omega}} \cos{\theta_{\omega}} \sin{\phi} \big) E(\omega)^2,  \notag \\
    P_z(2\omega) &= \varepsilon_0 \big( \chi_{zxx} \cos^{2}{\theta_{\omega}} + \chi_{zxx} \sin^{2}{\theta_{\omega}} \cos^{2}{\phi}  \notag \\
    &\quad + \chi_{zzz} \sin^{2}{\theta_{\omega}} \sin^{2}{\phi} \big) E(\omega)^2.
    \label{eq:nonlinear_polarization}
\end{align}
Here, $\theta_{\omega}$ and $\theta_{2\omega}$ are the polarization angles of the incident light and the analyzer, and $\phi$ is the incident angle (see Fig.~7).

For an analyzer angle parallel to the incident polarization $(\theta_\omega = \theta_{2\omega})$, the SHG intensity is given by:
\begin{multline}
    I^\mathrm{SHG}(2\omega) \propto \biggl| \chi_{yyy} \cos{3 \theta_{\omega}} 
    + \left( S_{1, 1} \sin{\theta_{\omega}} + S_{1, 3} \sin{3 \theta_{\omega}} \right) \sin{\phi}  \\
    + \frac{1}{4} \chi_{yyy} \left( \cos{\theta_{\omega}} -3 \cos{3 \theta_{\omega}} \right) \sin^{2}{\phi}  \\
    + \left( S_{3, 1} \sin{\theta_{\omega}} + S_{3, 3} \sin{3 \theta_{\omega}} \right) \sin^{3}{\phi} \biggl|^2 I^\mathrm{Laser}(\omega)^2
    \label{eq:shg_intensity}
\end{multline}
where
\begin{align*}
	S_{1, 1} &= \frac{1}{4} \left( -5 \chi_{xxz} + 4 \chi_{zxx} \right), \\	S_{1, 3} &= \frac{1}{4} \chi_{xxz}, \\
	S_{3, 1} &= \frac{3}{4} \left( \chi_{xxz} - \chi_{zxx} + \chi_{zzz} \right), \\ S_{3, 3} &= \frac{1}{4} \left( - \chi_{xxz} + \chi_{zxx} - \chi_{zzz} \right).
\end{align*}
%We assume real values for $\chi_{ijk}^{(2)}$ for simplicity.

To determine the in-plane crystal orientation, we performed polarization-resolved SHG measurements (Fig.~1(d)).
The SHG intensity was recorded while rotating the incident polarization angle $\theta_\omega$ and analyzer angle $\theta_{2\omega}$ simultaneously over $360^\circ$.
This produced a six-lobe pattern at angles $\phi = 0^\circ, 60^\circ, 120^\circ, 180^\circ, 240^\circ, 300^\circ$. 
The measured pattern aligns well with Eq.~(\ref{eq:shg_intensity}), confirming the expected $3m$ symmetry~\cite{Sekine_2022_APL}. These results establish SHG as a precise, non-destructive method for determining the crystal orientation in the van der Waals material trigonal PtBi$_{2}$.

\section{appendix B: Details of the magnetic field dependence and temperature dependence of the nonreciprocity under in-plane magnetic field}

\subsection{1. Details of the magnetic field dependence}

\begin{figure}
\includegraphics[width=85mm]{theta_dep.png}
\caption{\label{fig:epsart} (a) Schematic illustration of the measurement configuration under an in-plane magnetic field with a small misalignment of the magnetic field direction toward the out-of-plane direction $\delta_{\rm 1}$. (b), (c) Magnetic field dependence of (b) $R^{\rm 2\omega, peak}_{\rm sum}$ (red) and $\Delta I_{\rm c, sum}$ (blue), and (c) $R^{\rm 2\omega, peak}_{\rm diff}$ (red) and $\Delta I_{\rm c, diff}$ (blue) at 50~mK for the $\theta = 0 + \delta_{\rm 1} (>0)$ configuration. 
(d) Schematic illustration of the measurement configuration under an in-plane magnetic field with a small misalignment of the magnetic field direction away from the out-of-plane direction $\delta_{\rm 2}$. (e), (f) Magnetic field dependence of (e) $R^{\rm 2\omega, peak}_{\rm sum}$ (red) and $\Delta I_{\rm c, sum}$ (blue), and (f) $R^{\rm 2\omega, peak}_{\rm diff}$ (red) and $\Delta I_{\rm c, diff}$ (blue) at 50~mK for the $\theta = 0 -\delta_{\rm 2} (<0)$ configuration.
}
\end{figure}

To further evaluate the hysteretic components of the nonreciprocity, we calculate the summation and difference of the data between $\mu_0 H_{\rm p} = 1$~T and $\mu_0 H_{\rm p} = -1$~T. The summation and difference of $R^{\rm 2\omega, peak}$ and $\Delta I_{\rm c}$ are defined as follows:
\begin{align}
    R^{\rm 2\omega, peak}_{\rm sum} &= \frac{1}{2} \left\{ R^{\rm 2\omega, peak}(\mu_0 H_{\rm p} = 1~\text{T}) \right. \nonumber \\
    &\qquad \left. + R^{\rm 2\omega, peak}(\mu_0 H_{\rm p} = -1~\text{T}) \right\}, \\
    R^{\rm 2\omega, peak}_{\rm diff} &= \frac{1}{2} \left\{ R^{\rm 2\omega, peak}(\mu_0 H_{\rm p} = 1~\text{T}) \right. \nonumber \\
    &\qquad \left. - R^{\rm 2\omega, peak}(\mu_0 H_{\rm p} = -1~\text{T}) \right\}, \\
    \Delta I_{\rm c, sum} &= \frac{1}{2} \left\{ \Delta I_{\rm c}(\mu_0 H_{\rm p} = 1~\text{T}) \right. \nonumber \\
    &\qquad \left. + \Delta I_{\rm c}(\mu_0 H_{\rm p} = -1~\text{T}) \right\}, \\
    \Delta I_{\rm c, diff} &= \frac{1}{2} \left\{ \Delta I_{\rm c}(\mu_0 H_{\rm p} = 1~\text{T}) \right. \nonumber \\
    &\qquad \left. - \Delta I_{\rm c}(\mu_0 H_{\rm p} = -1~\text{T}) \right\}.
\end{align}
These quantities are shown in Figs.~8(b) and 8(c). 
The summation of the nonreciprocal transport properties, which excludes the hysteretic components, exhibits an odd function with respect to the magnetic field, as shown in Fig.~8(b). In contrast, the difference of the nonreciprocal transport properties, representing the hysteretic components, exhibits an even function, as shown in Fig.~8(c). 

\subsection{2. Effect of misalignment of magnetic field direction}

To understand the behavior of the in-plane magnetic field dependence of nonreciprocal transport properties, we discuss the effect of the misalignment of the magnetic field direction with respect to the out-of-plane direction $\theta$ (see Fig.~8(a)). In this case, we define $\theta$ as positive. In our experimental setup, it is difficult to align the in-plane magnetic field perfectly perpendicular both to the current direction and the $c$-axis; some misalignment toward the out-of-plane direction is inevitable. 
Thus, we intentionally tilted the device holder toward the negative side (i.e., $\theta<0$)
with respect to the $c$-axis at room temperature and cooled the device down to 50~mK, 
as shown in Fig.~8(d). 
Figures~8(e) and 8(f) present the same measurements as Figs.~8(b) and 8(c), but with the opposite sign of the misalignment toward the out-of-plane direction. 
The summation of the nonreciprocal transport properties, which excludes the hysteretic components, changes its sign when the misalignment direction is reversed (see Figs.~8(b) and 8(e)). In contrast, the difference of the nonreciprocal transport properties, which represents the hysteretic components, remains unchanged, particularly around zero magnetic field (see Figs.~8(c) and 8(f)).

These results suggest that the summation of the nonreciprocal transport properties (Figs.~8(b) and 8(e)), which excludes the hysteretic components, originates from the misaligned magnetic field component parallel to the out-of-plane direction. This is supported by comparison with the nonreciprocal transport under an out-of-plane magnetic field. As shown in Fig.~5(i), the nonreciprocity under the out-of-plane magnetic field exhibits an odd function with respect to the magnetic field, similar to the behavior of the summation of the nonreciprocal transport properties under an in-plane magnetic field. Furthermore, the nonreciprocity under the out-of-plane magnetic field does not show hysteretic behavior with respect to $H_{\rm p}$, as demonstrated in Appendix~E, which is consistent with the summation of the nonreciprocal transport properties under an in-plane magnetic field.
In contrast, the difference of the nonreciprocal transport properties (Figs.~8(c) and 8(f)), which represents the hysteretic components, is attributed to the in-plane magnetic field. These findings indicate that the $H_{\rm p}$-dependent zero-field nonreciprocal transport properties arise from the in-plane magnetic field.

\subsection{3. Temperature dependence}

\begin{figure}
\includegraphics[width=50mm]{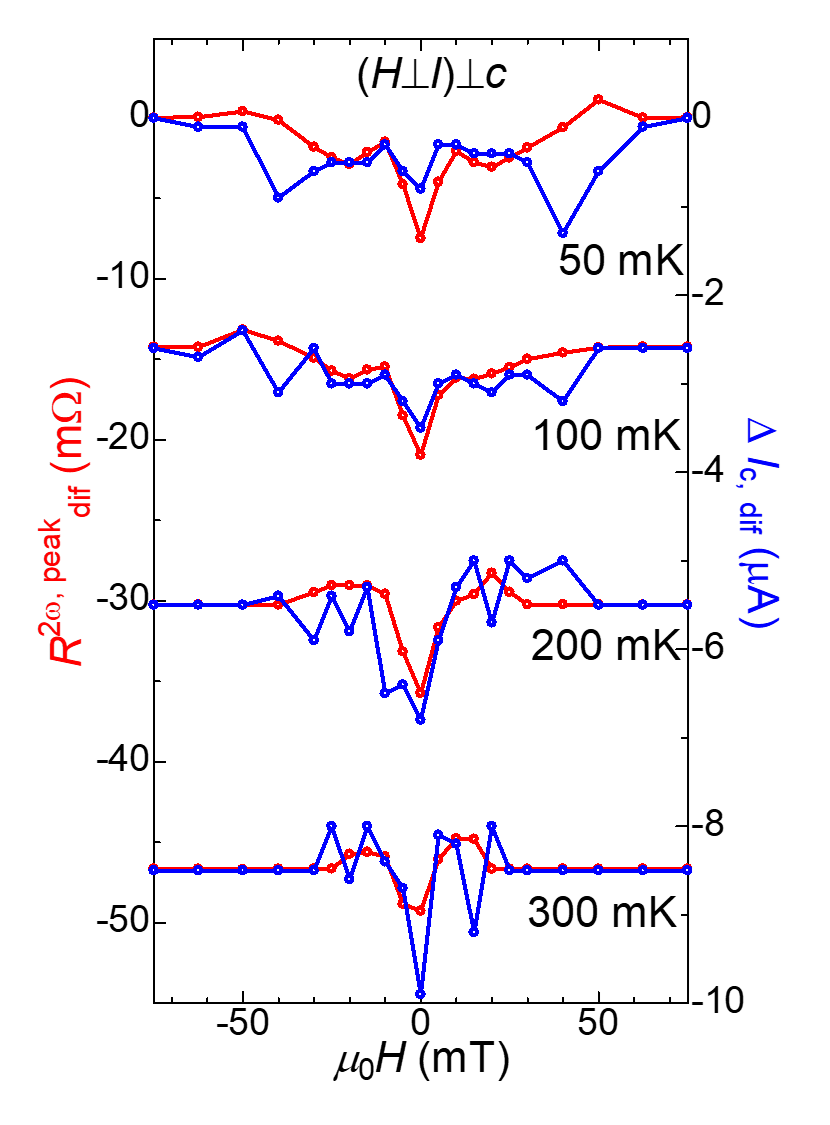}
\caption{\label{fig:epsart} The magnetic field dependence of $R^{\rm 2\omega, peak}_{\rm dif}$ (red) and $\Delta I_{\rm c, dif}$ (blue) at various temperatures under a magnetic field perpendicular to the current for $|\mu_0 H_{\rm p}| = 1$~T.
}
\end{figure}

We investigate the temperature dependence of the nonreciprocal transport properties. Figure~9 shows the magnetic field dependence of $R^{\rm 2\omega, peak}_{\rm dif}$ (red) and $\Delta I_{\rm c, dif}$ (blue) at various temperatures under an in-plane magnetic field perpendicular to the current for $|\mu_0 H_{\rm p}| = 1$~T. The nonreciprocity decreases monotonically with increasing temperature.

\section{appendix C: nonreciprocal transport properties for poling field 2~T}

\begin{figure}
\includegraphics[width=85mm]{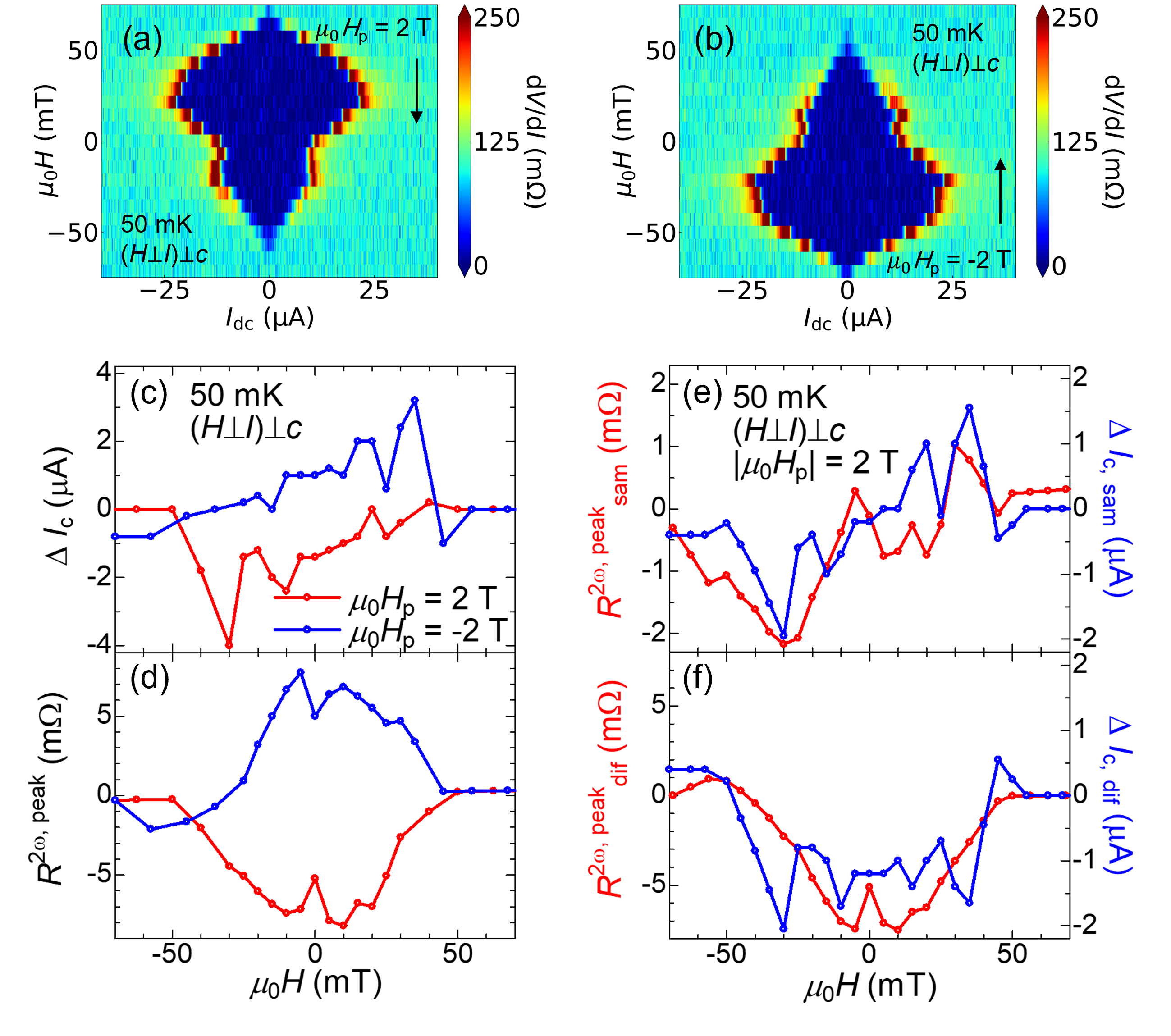}
\caption{\label{fig:epsart} (a, b) $\text{d}V/\text{d}I$ as a function of $I_{\text{dc}}$ and $\mu_0 H$ at 50~mK, measured after applying (a) $\mu_0 H_{\rm p} = 2$~T and (b) $\mu_0 H_{\rm p} = -2$~T under an in-plane magnetic field perpendicular to the current. 
(c, d) Magnetic field dependence of (c) $\Delta I_{\rm c}$ and (d) $R^{\rm 2\omega, peak}$ at 50~mK for $\mu_0 H_{\rm p} = 2$~T (red) and $\mu_0 H_{\rm p} = -2$~T (blue) under an in-plane magnetic field perpendicular to the current. 
(e, f) Magnetic field dependence of (e) $R^{\rm 2\omega, peak}_{\rm sum}$ (red) and $\Delta I_{\rm c, sum}$ (blue), and (f) $R^{\rm 2\omega, peak}_{\rm diff}$ (red) and $\Delta I_{\rm c, diff}$ (blue) at 50~mK under an in-plane magnetic field perpendicular to the current for $|\mu_0 H_{\rm p}| = 2$~T.
}
\end{figure}

To examine the $\mu_0 H_{\rm p}$ dependence of the hysteresis, we present the transport properties for $|\mu_0 H_{\rm p}| = 2$~T in Fig.~10. Figures~10(a) and 10(b) show $\text{d}V/\text{d}I$ as a function of $I_{\text{dc}}$ and $\mu_0 H$ at 50~mK, measured after applying $\mu_0 H_{\rm p} = \pm2$~T under an in-plane magnetic field perpendicular to the current. Compared to $|\mu_0 H_{\rm p}| = 1$~T (Figs.~3(e) and 3(f)), the zero-resistance region exhibits greater distortion.

Figures~10(c) and 10(d) show the magnetic field dependence of $\Delta I_{\rm c}$ and $R^{\rm 2\omega, peak}$ at 50~mK for $\mu_0 H_{\rm p} = \pm2$~T. Similar to the analysis for $|\mu_0 H_{\rm p}| = 1$~T, we calculate the summation and difference of the data between $\mu_0 H_{\rm p} = 2$~T and $\mu_0 H_{\rm p} = -2$~T and plot them in Figs.~10(e) and 10(f), showing $R^{\rm 2\omega, peak}_{\rm sum}$ (red) and $\Delta I_{\rm c, sum}$ (blue), as well as $R^{\rm 2\omega, peak}_{\rm diff}$ (red) and $\Delta I_{\rm c, diff}$ (blue).

While the nonreciprocity differs slightly from that at $|\mu_0 H_{\rm p}| = 1$~T (see Figs.~3 and 8), both even and odd components with respect to the magnetic field persist. This variation indicates that the trapped magnetic flux changes with $H_{\rm p}$, further confirming that the zero-field nonreciprocal transport phenomena originate from these trapped magnetic flux.

\section{appendix D: external magnetic field correction}

\begin{figure}
\includegraphics[width=85mm]{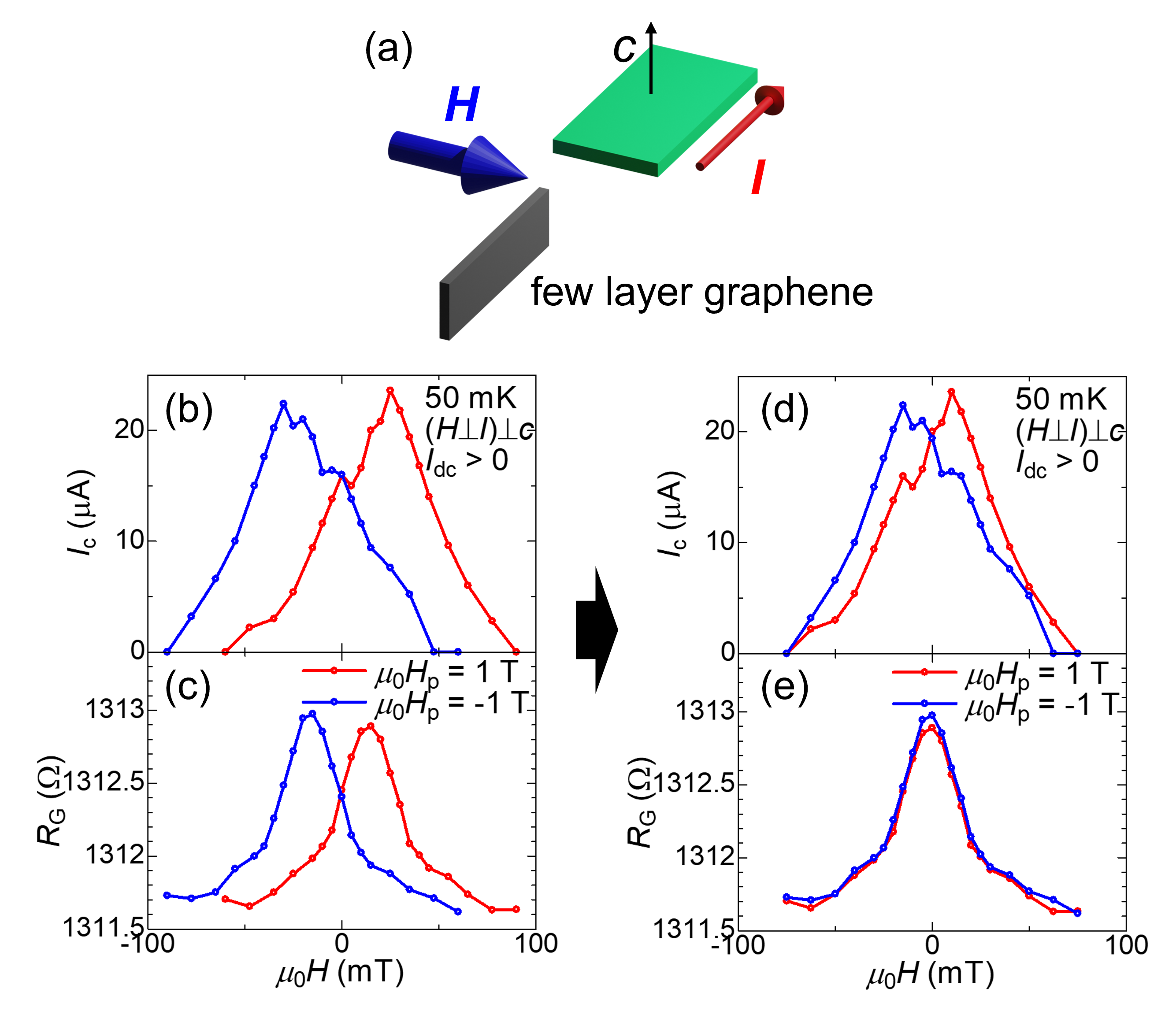}
\caption{\label{fig:epsart} (a) Schematic illustration of the measurement configuration with few-layer graphene. A magnetic field was applied perpendicular to the plane of the few-layer graphene to determine the zero magnetic field. (b, c) Raw data showing the magnetic field dependence of (b) the critical current of trigonal PtBi$_2$ ($I_{\rm c}$) and (c) the resistance of the few-layer graphene ($R_{\rm G}$) at 50~mK for $\mu_0 H_{\rm p} = 1$~T (red) and $\mu_0 H_{\rm p} = -1$~T (blue). $I_{\rm c}$ was measured for positive DC current under a magnetic field applied in-plane and perpendicular to the current. 
(d, e) Magnetic field dependence of (d) $I_{\rm c}$ and (e) $R_{\rm G}$ after correcting the zero magnetic field by shifting the data, assuming that $R_{\rm G}$ should be an even function of the magnetic field.
}
\end{figure}

Because a superconducting magnet was used in this study, hysteresis due to the magnet should be observed at low magnetic fields after applying a large magnetic field. To ensure the accuracy of the external magnetic field, the resistance of the few-layer graphene ($R_{\rm G}$) was simultaneously measured along with the transport properties of trigonal PtBi$_2$, as illustrated in Fig.~11(a). 
Figures~11(b) and 11(c) show the raw data for the magnetic field dependence of the critical current of trigonal PtBi$_2$ ($I_{\rm c}$) and the resistance of the few-layer graphene ($R_{\rm G}$) at 50~mK for $\mu_0 H_{\rm p} = 1$~T (red) and $\mu_0 H_{\rm p} = -1$~T (blue), respectively. $I_{\rm c}$ was measured for positive DC current under a magnetic field applied in-plane and perpendicular to the current. Hysteresis with respect to $\mu_0 H_{\rm p}$ is observed in both $I_{\rm c}$ and $R_{\rm G}$. 
Since $R_{\rm G}$ should not exhibit hysteresis and is expected to be an even function of the magnetic field, the hysteresis in $R_{\rm G}$ is attributed to the hysteresis of the superconducting magnet. To correct experimental errors caused by the hysteresis of superconducting magnet, the magnetic field values of $\approx \pm$15~mT were shifted to align the magnetic field dependence of $R_{\rm G}$ for $\mu_0 H_{\rm p} = 1$~T and $\mu_0 H_{\rm p} = -1$~T. The corrected data are shown in Figs.~11(d) and 11(e). 
This correction was applied to all data presented in this paper except for the raw data shown in Figs.~11(b) and 11(c).

\section{appendix E: hysteresis under an out-of-plane magnetic field configuration}

\begin{figure}
\includegraphics[width=50mm]{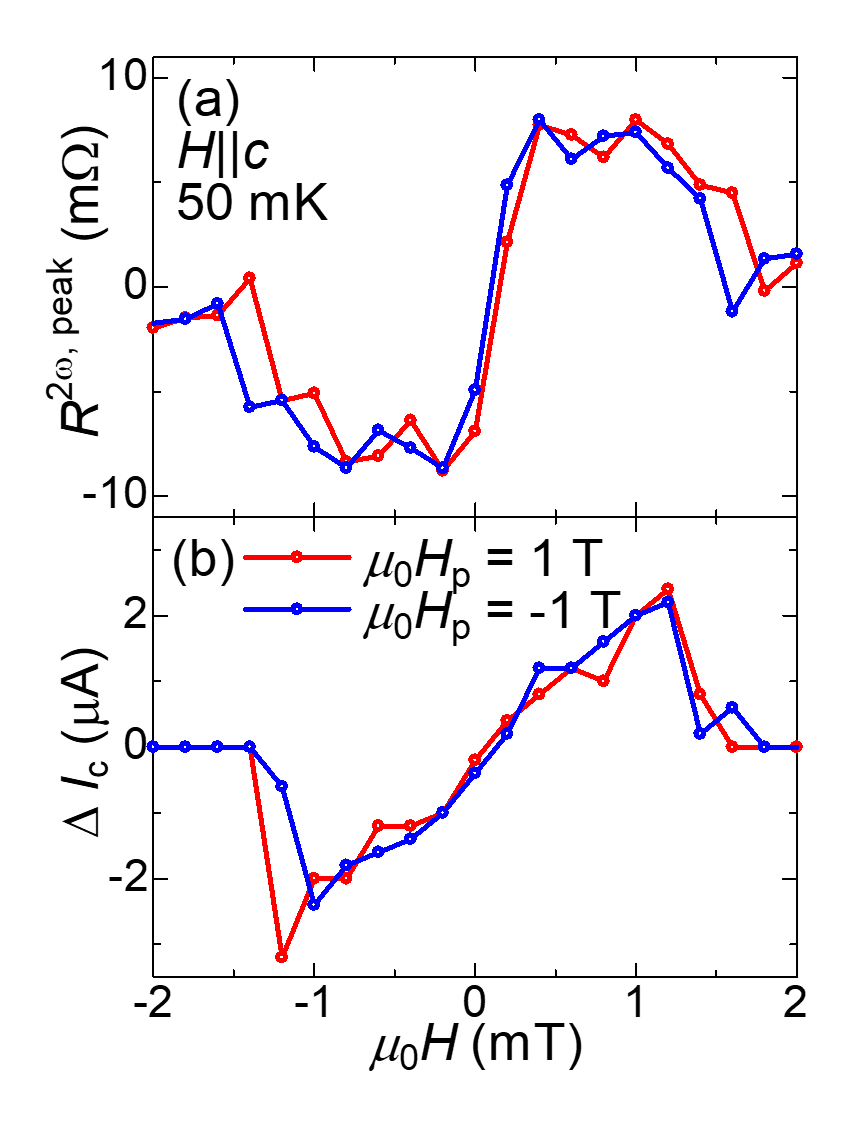}
\caption{\label{fig:epsart} (a, b) Out-of-plane magnetic field dependence of (a) $\Delta I_{\rm c}$ and (b) $R^{\rm 2\omega, peak}$ at 50~mK for $\mu_0 H_{\rm p} = 1$~T (red) and $\mu_0 H_{\rm p} = -1$~T (blue).
}
\end{figure}

Because hysteresis with respect to $\mu_0 H_{\rm p}$ is observed under an in-plane magnetic field, we also investigated its effect under an out-of-plane magnetic field. Figure~12(a) and 12(b) shows the out-of-plane magnetic field dependence of $R^{\rm 2\omega, peak}$ and $\Delta I_{\rm c}$, respectively, measured at $T=50$~mK with $\mu_0 H_{\rm p} = 1$~T (red) and $\mu_0 H_{\rm p} = -1$~T (blue). No hysteresis is observed in the out-of-plane magnetic field configuration.
These results indicate that magnetic flux in trigonal PtBi$_2$ are trapped when an in-plane magnetic field is applied but not for an out-of-plane magnetic field. Furthermore, the absence of zero-field nonreciprocity under an out-of-plane magnetic field confirms that the observed zero-field nonreciprocity under an in-plane magnetic field is intrinsic and not due to experimental errors.

\section{appendix F: reproducibility}

\begin{figure}
\includegraphics[width=85mm]{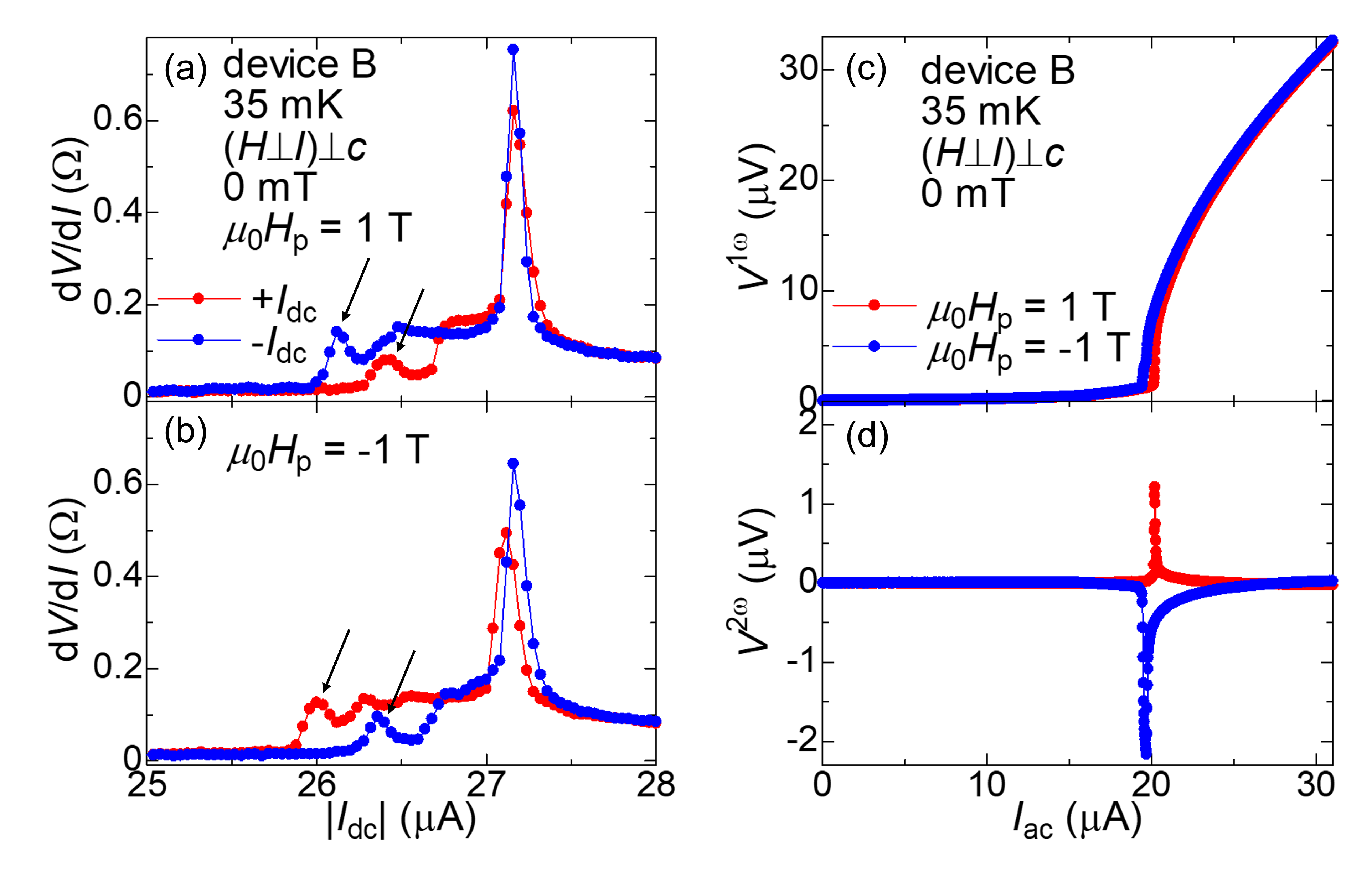}
\caption{\label{fig:epsart} (a, b) $I_{\text{dc}}$ dependence of $\text{d}V/\text{d}I$ for device B (thickness $45~\mathrm{nm}$) at 0~mT and 50~mK for (a) $\mu_0 H_{\rm p} = 1$~T and (b) $\mu_0 H_{\rm p} = -1$~T. Red and blue curves correspond to opposite $I_{\text{dc}}$ directions. 
(c, d) $I_{\text{ac}}$ dependence of (c) $V^{1\omega}$ and (d) $V^{2\omega}$ for device B at 0~mT and 50~mK. Red and blue curves represent data for $\mu_0 H_{\rm p} = 1$~T and $\mu_0 H_{\rm p} = -1$~T, respectively.
}
\end{figure}

\begin{figure}
\includegraphics[width=50mm]{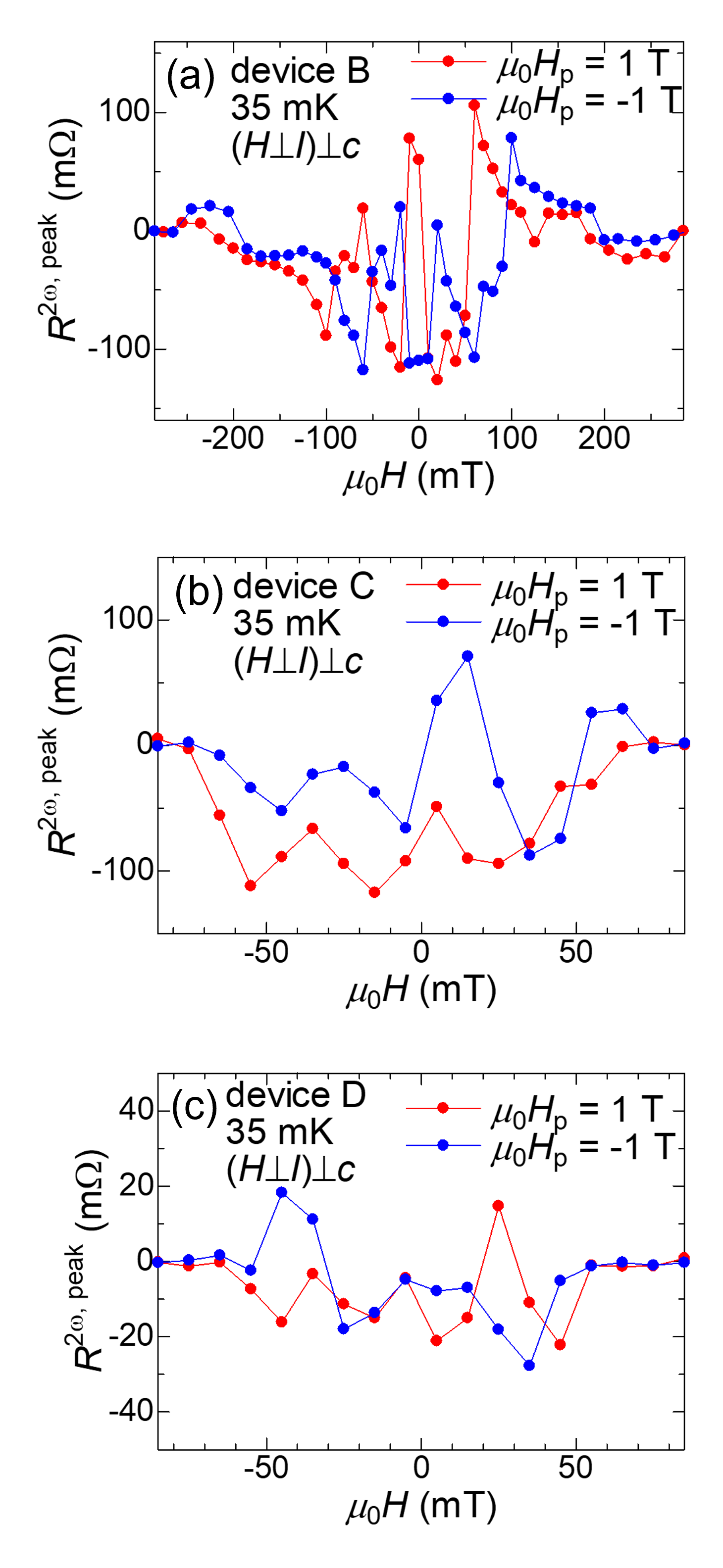}
\caption{\label{fig:epsart} Magnetic field dependence of $R^{\rm 2\omega, peak}$ at 50~mK under an in-plane magnetic field perpendicular to the current for $\mu_0 H_{\rm p} = 1$~T (red) and $\mu_0 H_{\rm p} = -1$~T (blue) for (a) device B, (b) device C (thickness $65~\mathrm{nm}$), and (c) device D (thickness $80~\mathrm{nm}$).
}
\end{figure}

To verify the reproducibility of the zero-field nonreciprocal transport, we present data for other devices in Figs.~13 and 14. Figures~13(a) and 13(b) show the $I_{\text{dc}}$ dependence of $\text{d}V/\text{d}I$ obtained with device B at $\mu_{0}H=0$~mT and $T=50$~mK for $\mu_0 H_{\rm p} = 1$~T and $\mu_0 H_{\rm p} = -1$~T, respectively. Red and blue curves correspond to opposite $I_{\text{dc}}$ directions. Although there are at least two peaks, indicating the presence of superconducting domains with different critical currents, the first peak depends on the $I_{\text{dc}}$ direction. This superconducting diode effect in device B also changes its sign depending on the sign of $H_{\rm p}$.
Figures~13(c) and 13(d) show the $I_{\text{ac}}$ dependence of $V^{1\omega}$ and $V^{2\omega}$, respectively, at $\mu_{0}H=0$~mT and $T=50$~mK. Red and blue curves represent data for $\mu_0 H_{\rm p} = 1$~T and $\mu_0 H_{\rm p} = -1$~T, respectively. $V^{2\omega}$ appears near the critical current and also changes its sign depending on the sign of $H_{\rm p}$.

In Figs.~14(a)-14(c), we show the magnetic field dependence of $R^{\rm 2\omega, peak}$ at $T=50$~mK under a magnetic field perpendicular to the current for $\mu_0 H_{\rm p} = 1$~T (red) and $\mu_0 H_{\rm p} = -1$~T (blue), measured with device B, device C, and device D, respectively. Although the magnetic field dependence appears somewhat scattered, the nonreciprocity exhibits hysteresis with respect to $H_{\rm p}$. These results confirm the magnetic flux-induced nonreciprocity discussed in the main text.
It should be noted that these devices were prepared using the same scotch tape on the same substrate, meaning that they share the same polarity $P$. However, despite having the same $P$, device B exhibits an opposite sign of zero-field nonreciprocity compared to devices C and D. This result indicates that the origin of the zero-field nonreciprocity in trigonal PtBi$_2$ is not solely determined by the polarity $P$.

%\bibliography{ptbi2_citation}% Produces the bibliography via BibTeX.

%

\end{document}